\documentclass[prd,preprint,showpacs,byrevtex]{revtex4}

\newcommand{\be}{\begin{equation}}
\newcommand{\ee}{\end{equation}}
\newcommand{\bea}{\begin{eqnarray}}
\newcommand{\eea}{\end{eqnarray}}
\usepackage[dvips]{graphicx}
\begin{document}
\draft
%
%
\title{Minijet Initial Conditions For Non-Equilibrium Parton
Evolution at RHIC and LHC}
\author{Fred Cooper, Emil Mottola and Gouranga C. Nayak}

\address
{\small\it{T-8, Theoretical Division, Los Alamos National Laboratory,
Los Alamos, NM 87545, USA }}
     

\begin{abstract}
An important ingredient for the non-equilibrium evolution of partons at RHIC
and LHC is to have some physically reasonable 
initial conditions for the single particle phase
space distribution functions for the partons. We consider several 
plausible 
parametrizations of initial conditions for the single particle
distribution function $f_i(x,p)$ and fix the parameters by matching  
$\int f(x,p)p^\mu d \sigma_\mu$ to
the invariant  momentum space semi-hard parton distributions obtained using
perturbative QCD (pQCD), as well as fitting low order moments of the 
distribution function. 
We consider parametrizations of $f_i(x,p)$  with both boost 
invariant and boost non-invariant assumptions. 
We determine the  initial 
number density, energy density and the corresponding (effective)
temperature of the minijet plasma at RHIC and LHC energies. 
For a boost non-invariant minijet phase-space distribution
function we obtain $\sim$ 30(140) /fm$^3$ as the initial number density,
$\sim$ 50(520) GeV/fm$^3$ as the initial energy density and
$\sim$ 520(930) MeV as the corresponding initial effective 
temperature at RHIC(LHC). 
\end{abstract}  
\bigskip

\maketitle    
%

In order to study thermalization of quark-gluon plasma 
at RHIC and LHC in any kinetic approach,
one needs suitable initial conditions for the single 
particle phase-space distribution of partons.
The hard and semi hard
parton production can be calculated by using perturbative QCD (pQCD) 
\cite{hijing}. The soft
parton production cannot be calculated in a kinetic approach
within perturbative QCD. 
One approximate strategy for looking at soft parton production is to assume the 
creation of a classical chromofield \cite{larry1,others,nus}. 
The first problem one faces in specifying
initial conditions
for the phase space single particle distribution function $f(x,p,t_0)$
of partons is that
from pQCD one has information only about the momentum space 
distribution of minijets. The pQCD calculation
(using scattering matrix element squared
and the structure functions) is done in an approach which calculates 
probabilities between initial and final states so that any information
about space-time evolution is lost. Thus we need to supplement pQCD
with assumptions about the dependence of $f(x,p)$ on $x$.

In this paper 
our strategy will be to use pQCD to determine the single particle distribution
function for the partons in momentum space, and then by taking simple
parametric forms for $f(x,p)$ constrain 
the parametrization by fitting the single particle distribution
as well as the average transverse momentum.
This process can be refined by adding more parameters and fitting further 
moments
of the jet distribution function. Of course, one hopes that the details of the
parametrization of the initial conditions do not significantly effect the thermalization
process.  

The lowest order pQCD inclusive ($2 \rightarrow 2$) minijet cross
section per nucleon in A-A collision is given by: 
\begin{equation}
\sigma_{jet} = \int dp_T dy_1 dy_2 {{2 \pi p_T} \over {\hat{s}}} 
\sum_{ijkl}
x_1~ f_{i/A}(x_1, p_T^2)~ x_2~ f_{j/A}(x_2, p_T^2)~
\hat{\sigma}_{ij \rightarrow kl}(\hat{s}, \hat{t}, \hat{u}),
\label{jet}
\end{equation}
where $\hat{\sigma}_{ij \rightarrow kl}$ is the elementary pQCD parton cross
sections for the process $ij \rightarrow kl$. As gluons are the dominant
part of the total minijet production, we will only consider the processes
$gq(\bar q) \rightarrow gq(\bar q)$ and $gg \rightarrow gg$ in this paper.
The partonic level cross sections for these processes are given by:
\begin{equation}
\hat{\sigma}_{gq \rightarrow gq} = {{\alpha_s^2} \over
{ \hat{s}}} ({\hat{s}^2+\hat{u}^2}) [ 
{{1} \over {\hat{t}^2}}
- {{4} \over {9\hat{s}\hat{u}}}], 
\nonumber
\end{equation}
\begin{equation}
\hat{\sigma}_{gg \rightarrow gg} = {{9 \alpha_s^2} \over
{2 \hat{s}}} [ 3 - {{\hat{u}\hat{t}} \over {\hat{s}^2}}
 - {{\hat{u}\hat{s}} \over {\hat{t}^2}}
 - {{\hat{s}\hat{t}} \over {\hat{u}^2}}].
\nonumber
\end{equation}
The rapidities $y_1$, $y_2$ and the momentum fractions $x_1$, $x_2$ are
related by,
\begin{equation}
x_1=p_T~(e^{y_1}+e^{y_2})/{\sqrt{s}}, \hspace{0.5cm}
x_2=p_T~(e^{-y_1}+e^{-y_2})/{\sqrt{s}},
\nonumber
\end{equation}
with the kinematic relations:
\begin{equation}
\hat{s}=x_1 x_2 s, ~~~~~~{\rm{and}}~~~~~~\hat t=-\frac{\hat s}{2}[1-\tanh (\frac{y_1-y_2}{2})].
\end{equation}
To compute the minijet cross section using Eq. (\ref{jet}) we need to 
specify the minimum transverse momentum cut-off $p_0$ above which the 
incoherent parton
picture is applicable. The momentum cut-off is $\sim$ 1 GeV at RHIC
and $\sim$ 2 GeV at LHC, which are obtained from the argument that there is
saturation of the gluon structure function 
at low $x$ \cite{al}. For our
quantitative calculation we use $p_0$= 1 GeV at RHIC and 2 GeV at LHC.
The nuclear modified parton distribution functions which includes the shadowing
effects are given by $f_{i/A}(x,Q^2)=R_A(x,Q^2)f_{i/N}(x,Q^2)$ where
$A$ and $N$ stand for nucleus and free nucleon respectively.
In this paper we use the GRV98 set of parton distributions for the free
nucleon distribution function
$f_{i/N}(x,Q^2)$ \cite{grv98} (in momentum space). The GRV98 analysis
uses the low $x$ HERA data on deep inelatstic scatterings along with
data from other hard scattering processes at fixed $Q$. The $Q^2$ evolution
of the parton distribution function is performed by using
perturbative QCD evolution equations.
We use the EKS98 numerical
parametrizations for the ratio function $R_A(x,Q^2)$ \cite{eks98}.
The EKS98 parametrization uses NMC and E665 structure functions
from deep inelastic lepton-nucleus
collisions and E772 Drell-Yan data from proton-nucleus collisions
at fixed $Q$. The $Q^2$ evolution of the structure function is
studied by using the DGLAP evolution equation.
We multiply the above minijet cross section by a standard 
K factor (K=2), to account for the higher order contributions.

For central collisions, the minijet cross section (Eq.\ (\ref{jet}))
can be related to the total number of partons $(N^{jet})$ by
\begin{equation}
\frac{dN^{jet}}{dydp_T}=K~T(0) ~ \int dy_2 {{2 \pi p_T} \over {\hat{s}}} 
\sum_{ijkl} x_1~ f_{i/A}(x_1, p_T^2)~ x_2~ f_{j/A}(x_2, p_T^2)~
\hat{\sigma}_{ij \rightarrow kl}(\hat{s}, \hat{t}, \hat{u}),
\label{number}
\end{equation}
where $T(0)= 9A^2/{8\pi R_A^2}$ fm is the nuclear geometrical factor for
head-on AA collisions (for a nucleus with a sharp surface).
Here $R_A=1.1 A^{1/3}$ is the nuclear radius. 
Similarly the transverse energy distribution of the minijets are given by:
\begin{equation}
\frac{dE_T^{jet}}{dydp_T}~=~K~
T(0) ~ \int dy_2 {{2 \pi p_T^2} \over {\hat{s}}} 
\sum_{ijkl} x_1~ f_{i/A}(x_1, p_T^2)~ x_2~ f_{j/A}(x_2, p_T^2)~ 
\hat{\sigma}_{ij \rightarrow kl}(\hat{s}, \hat{t}, \hat{u}).
\label{energy}
\end{equation}
Our numerical results for the $p_T$ distribution of the minijets computed
from eq. (\ref{number}) are plotted in Fig. 1.
\begin{figure}
   \centering
   \includegraphics{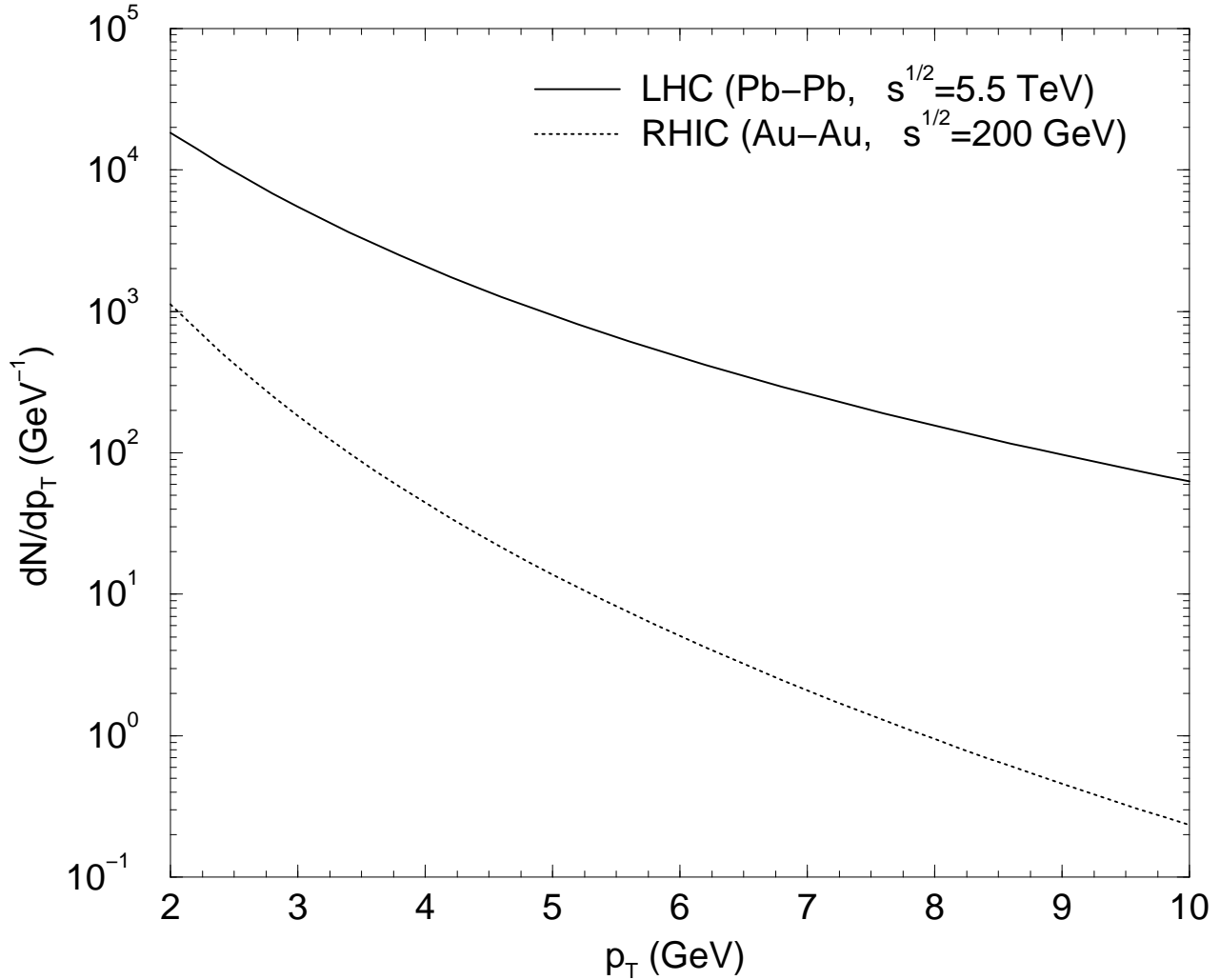}
   \caption{Transverse momentum 
distribution of the initial minijets at RHIC and LHC}
\label{dndpt}
\end{figure}

The rapidity distributions of the minijet production at RHIC and
LHC are obtained by integrating over $p_T$ for $p_T > p_0$ are
shown in Fig. 2. It can be seen that the minijet distribution
is not flat over the whole rapidity range. 
\begin{figure}
   \centering
   \includegraphics{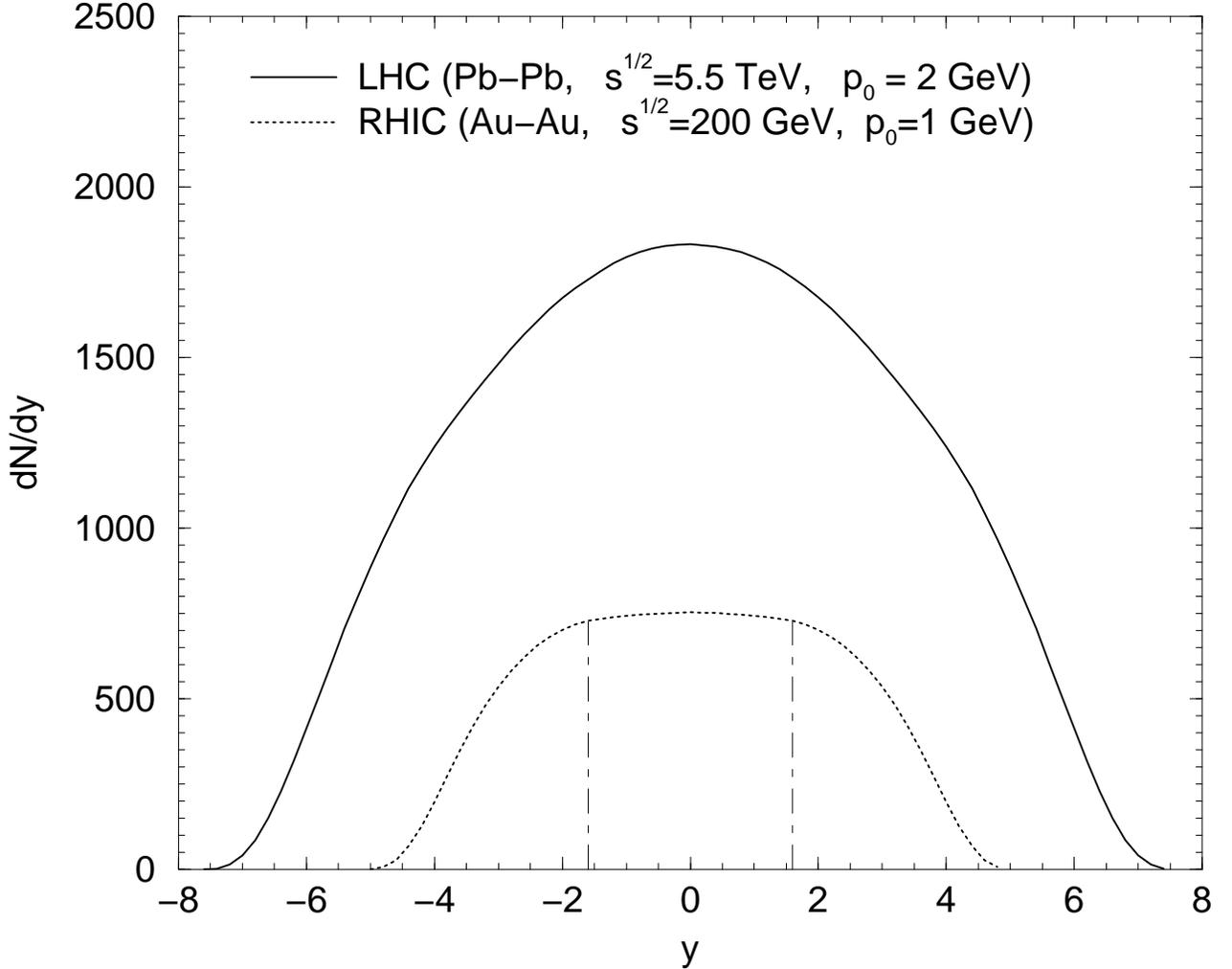}
   \caption{Rapidity distribution of the initial minijets at RHIC and LHC}
\label{temp}
\end{figure}
Hence it is not a particularly good approximation to assume boost invariance
\cite{bjorken} and restrict oneself to the  mid rapidity region. 
One important question that the transport equations
should answer is how different the final rapidity distribution 
of the hadrons (or other signatures) from the rapidity distribution
of the minijets at the pre-equilibrium stage are \cite{geig}. 
Of course this needs to be
supplemented by further dynamical evolution near the hadronic phase transition.
For this reason it is not appropriate to take into account just
the flat minijet rapidity distribution seen in the mid rapidity region 
when evolving the minijet plasma. 
Partons formed outside the mid rapidity region
do propagate in the pre-equilibrium stage with their rescatterings
and hence one must consider all the partons
in determining the initial condition of the QGP in the pre-equilibrium stage. 

Thus we require additional
information about how these partons are distributed in coordinate space.
A simple ansatz based on scaling (\cite{bjorken}) allows
extracting energy density from the 
transverse energy $E_T$ distribution via:
\be
\epsilon = \frac{1}{\pi R_A^2 \tau_0} \frac{d E_T}{d y}
\label{edbj}
\ee
This formula is correct if partons are uniformly distributed in 
coordinate space and the momentum rapidity $y$ is equal to the coordinate 
rapidity $\eta$. For an 1+1 dimensional
expanding system coordinate rapidity $\eta$ is defined via:
$t=\tau \cosh \eta$ and $z=\tau \sinh \eta$.
 
Traditionally the above formula is widely applied to obtain
energy density from $E_T$ distribution of minijets. 
Although the above formula might be applicable in a situation where the minijet
distribution is flat in rapidity, it is precisely the scale breaking effects
that we hope to understand from our transport approach. 
As it can be seen from Fig. 2
the initial minijet distribution function is not flat in the entire
rapidity range and hence the above formula is not applicable in the
pre-equilibrium region.
Also it is not clear that partons are uniformly distributed 
at the initial time and just dividing by a volume (see eq. (\ref{edbj}))
to obtain the energy density from the total energy is a rather crude 
estimate of initial conditions. 
To get a better estimate for the initial conditions 
one should take into account the correlation between
coordinate rapidity $\eta$ and momentum rapidity $y$. 
We will assume this correlation
has a particular form, and then look at the sensitivity of the
global quantities to the parametrization. 

For an expanding system the minijet number distribution eq. (\ref{number})
can be related to the phase-space distribution function via \cite{fred}:

\be
\frac{d^3 N^{jet}}{\pi dy dp_T^2} = g_C
\int f(x, p) ~p^{\mu} d\sigma_{\mu}.
\ee
where $g_C=16$ is the product of spin and color degrees of freedom, 
\bea
d\sigma^{\mu}=\pi R_A^2 \tau d\eta (\cosh \eta,0,0, \sinh \eta), 
\eea
and
\bea
p^{\mu}= (p_T \cosh y, p_T \cos \phi, p_T \sin \phi, p_T \sinh y),
\eea
for an 1+1 dimension expanding system.
Using the above relations we get at the initial time $\tau_0$ (=1/$p_0$):
\be
\frac{d N^{jet}}{\pi dy dp_T^2} = g_C \pi R_A^2 \tau_0 \int d\eta
~p_T \cosh(\eta-y)~ f(p_T, \eta, y, \tau_0),
\label{nb}
\ee
where we assume a transverse isotropy at the early stage.
In the following we will obtain the initial phase-space distribution
function of the gluon minijets by using both boost invariant and 
boost non-invariant schemes.

\subsection{Boost Invariant Initial Distribution Function From Minijets}
First let us assume that a boost invariant description \cite{bjorken} is appropriate. 
Then the gluon distribution function depends only on the boost invariant parameters
$\tau$, $p_T$ and $\xi=\eta-y$. We will parametrize the $\eta-y$ correlation 
as Gaussian:
\be
f(p_T, \eta-y, \tau_0) = f(p_T) 
e^{-\frac{(\eta-y)^2}{\sigma^2(p_T)}},
\label{inds}
\ee
where $f(p_T)$ will be determined from eq. (\ref{nb})
by using eq. (\ref{number}) which is obtained by using pQCD.
Using the above equation in eq. (\ref{nb}) we get:
\be
\frac{d N^{jet}}{dp_T} = g_C 2 \pi^2 R_A^2 \tau_0 ~p_T^2 f(p_T)
\int dy \int_{-\infty}^{\infty} d\eta \cosh (\eta-y)
e^{-\frac{(\eta-y)^2}{\sigma^2(p_T)}}
\label{nbt}
\ee
which gives:
\be
\frac{d N^{jet}}{dp_T} = g_C
2 \pi^2 \sqrt{\pi} R_A^2 \tau_0 p_T^2 
f(p_T)\sigma(p_T) e^{\sigma^2(p_T)/4} \int dy
\ee
where the maximum allowed value of the momentum rapidity 
at RHIC and LHC depends on $p_T$ and $\sqrt s$.
Integrating over $y$ we get from the above equation:
\be
f(p_T) =
\frac{\frac{d N^{jet}}{dp_T}}{ g_C 2 \pi^2 \sqrt{\pi} R_A^2 \tau_0 p_T^2  
2\ln(\sqrt s /{2 p_T}+ \sqrt{s/{4 p_T^2}-1}) }
\frac{e^{-\sigma^2(p_T)/4}}{{\sigma(p_T)}},
\ee
which gives a boost invariant initial distribution function:
\be
f(p_T, \xi, \tau_0) =
\frac{\frac{d N^{jet}}{dp_T}}{ g_C 2 \pi^2 \sqrt{\pi} R_A^2 \tau_0 p_T^2  
2\ln(\sqrt s /{2 p_T}+ \sqrt{s/{4 p_T^2}-1}) }
\frac{e^{-\frac{\xi^2}{\sigma^2(p_T)}}}{\sigma(p_T) e^{\sigma^2(p_T)/4}}.
\label{f0b}
\ee

Using this boost invariant initial
phase-space distribution function of the gluon 
the initial minijet number density is given by:
\bea
n(\tau_0)= g_C \int d\Gamma p^{\mu}u_{\mu} f(p_T, \eta-y,\tau_0) 
=g_C \int d^2p_T p_T \int d\xi \cosh \xi f(p_T,\xi,\tau) \nonumber \\
= 2 \pi \int dp_T\frac{\frac{d N^{jet}}{dp_T}}{ 2 \pi^2 
R_A^2 \tau_0 
2\ln(\sqrt s /{2 p_T}+ \sqrt{s/{4 p_T^2}-1}) },
\label{ndb}
\eea
where $d\Gamma=\frac{d^3p}{p^0}$.
Similarly the initial energy density is given by:
\bea
\epsilon(\tau_0)= g_C \int d\Gamma (p^{\mu}u_{\mu})^2 f(p_T, \eta-y,\tau_0) 
= g_C \int d^2p_T p_T^2 \int d\xi \cosh^2 \xi f(p_T,\xi,\tau) \nonumber \\
= 2 \pi \int dp_Tp_T\frac{\frac{d N^{jet}}{dp_T}}{ 4 \pi^2 
R_A^2 \tau_0 
2\ln(\sqrt s /{2 p_T}+ \sqrt{s/{4 p_T^2}-1}) } 
(e^{-\sigma^2(p_T)/4}+ e^{3\sigma^2(p_T)/4}).
\label{edb}
\eea

\subsection{Boost Non-Invariant Initial Distribution Function From Minijets}
In the above we used a boost invariant distribution function $f(\tau, \xi, p_T)$
as the initial phase-space distribution function. However this is
not consistent with the actual rapidity distribution function obtained
from pQCD. To relax the boost invariant
assumption, we will introduce a function $f(p_T,y)$ instead of $f(p_T)$ in eq. (\ref{inds}). 
The form of the boost non-invariant distribution we take is:
\be
f(p_T, \eta, y, \tau_0) = f(p_T,y) 
e^{-\frac{(\eta-y)^2}{\sigma^2(p_T)}},
\label{inds1}
\ee
where $f(p_T,y)$ is obtained from 
$\frac{d N^{jet}}{dy dp_T}$. Note that $f(p_T)$ in eq. (\ref{inds}) 
is obtained from $\frac{d N^{jet}}{dp_T dy}$ after integrating over
$y$. However, in the above form (eq. (\ref{inds1})), the $y$ dependence
is more general which for small $\sigma^2$ is close to that 
obtained by using pQCD.
Using the above form in eq. (\ref{nb}) we get:
\be
\frac{d N^{jet}}{dy dp_T} = g_C 2 \pi^2 R_A^2 \tau_0 ~p_T^2 f(p_T,y)
\int_{-\infty}^{\infty} d\eta \cosh (\eta-y)
e^{-\frac{(\eta-y)^2}{\sigma^2(p_T)}}
\label{nbt1}
\ee
which gives:
\be
\frac{d N^{jet}}{dy dp_T} = g_C
2 \pi^2 \sqrt{\pi} R_A^2 \tau_0 p_T^2 
f(p_T,y)\sigma(p_T) e^{\sigma^2(p_T)/4}.
\ee
From the above equation we get:
\be
f(p_T,y) =
\frac{\frac{d N^{jet}}{dy dp_T}}{ g_C 2 \pi^2 
\sqrt{\pi} R_A^2 \tau_0 p_T^2  
\sigma(p_T) e^{\sigma^2(p_T)/4}}
\ee
which gives a boost non-invariant initial phase-space gluon 
distribution function:
\be
f(p_T, \eta, y, \tau_0) =
\frac{\frac{d N^{jet}}{dy dp_T}}{ g_C 2 \pi^2 \sqrt{\pi} R_A^2 
\tau_0 p_T^2 }
\frac{e^{-\frac{(\eta-y)^2}{\sigma^2(p_T)}}}{\sigma(p_T) 
e^{\sigma^2(p_T)/4}}.
\ee

Using the above boost non-invariant gluon distribution function
the initial minijet number density is given by:
\bea
n(\tau_0,\eta)=g_C \int d\Gamma p^{\mu}u_{\mu} f(p_T, \eta, y,\tau_0) 
=g_C \int d^2p_T p_T \int dy \cosh (\eta-y) f(p_T,\eta, y, 
\tau_0) \nonumber \\
= 2 \pi \int dp_T \int dy 
\frac{\frac{d N^{jet}}{dy dp_T}}{ 2 \pi^2 \sqrt{\pi} R_A^2 \tau_0 }
\frac{e^{-\frac{(\eta-y)^2}{\sigma^2(p_T)}}}{\sigma(p_T) 
e^{\sigma^2(p_T)/4}} \cosh(\eta- y),
\label{ndnb}
\eea
and the initial minijet energy density given by:
\bea
\epsilon(\tau_0,\eta)= g_C \int d\Gamma (p^{\mu}u_{\mu})^2 
f(p_T, \eta, y,\tau_0) 
=g_C \int d^2p_T p_T^2 \int dy \cosh^2 (\eta-y) f(p_T,\eta, y, 
\tau_0) \nonumber \\
= 2 \pi \int dp_T p_T \int dy 
\frac{\frac{d N^{jet}}{dy dp_T}}{ 2 \pi^2 \sqrt{\pi} R_A^2 \tau_0 }
\frac{e^{-\frac{(\eta-y)^2}{\sigma^2(p_T)}}}{\sigma(p_T) 
e^{\sigma^2(p_T)/4}} \cosh^2(\eta-y).  
\label{ednb}
\eea

\subsection{Initial Energy Density and Number Density at RHIC and LHC}

Using the above expressions for the minijet initial phase-space 
distribution functions
we can now predict the initial energy density and number density of the
minijet plasma at RHIC and LHC. 
As temperature can not be defined in a non-equilibrium
situation we define an {\it effective} temperature by:
\be
T_{eff}=[\frac{15}{8\pi^2} \epsilon]^{1/4},
\ee
with the non-equilibrium energy density obtained above.

Before we calculate the initial conditions we would like to find out
the values of the unknown parameter $\sigma^2$ appearing in the above
equation. We will fix this unknown parameter by equating the first $p_T$
moment of the distribution function with the pQCD predicted $E_T$
distribution as given by eq. (\ref{energy}). We have for the boost invariant
case:
\bea
\int dp_T \frac{dE_T^{jet}}{dp_T}=
2g_C\pi^2 R_A^2 \tau_0 \int
dp_T ~p_T^3 f(p_T) 
= \int dp_T 
\frac{p_T \frac{d N^{jet}}{dp_T}}{ \sqrt{\pi} 
2\ln(\sqrt s /{2 p_T}+ \sqrt{s/{4 p_T^2}-1}) 
\sigma(p_T) e^{\sigma^2(p_T)/4}},
\label{bn}
\eea
and for the boost non-invariant case:
\bea
\int dp_T \int dy \frac{dE_T^{jet}}{dy dp_T}=
2g_C\pi^2 R_A^2 \tau_0 \int
dp_T \int dy ~p_T^3 f(p_T,y) 
= \int dp_T \int dy \frac{p_T 
\frac{d N^{jet}}{dy dp_T}}{ \sqrt{\pi} \sigma(p_T) e^{\sigma^2(p_T)/4}}.
\label{nbn}
\eea
We determine the value of $\sigma^2$ by using the pQCD values of
$ \frac{dE_T^{jet}}{dy dp_T}$ and $ \frac{dE_T^{jet}}{dp_T}$
from eq. (\ref{energy}) in the left hand side of the above equations.
Assuming $\sigma^2$ independent of $p_T$ we find from the above
equations: $\sigma^2$ = 0.0034(0.0015) at RHIC(LHC)
for boost invariant case and $\sigma^2$ = 0.28(0.28) at RHIC(LHC)
for boost non-invariant case. Note that for boost non-invariant case
$\sigma^2$ is same at RHIC and LHC as the normalization factor
${ \sqrt{\pi} \sigma(p_T) e^{\sigma^2(p_T)/4}}$ 
in eq. (\ref{nbn}) is energy independent whereas the corresponding
normalization factor
${ \sqrt{\pi} 
2\ln(\sqrt s /{2 p_T}+ \sqrt{s/{4 p_T^2}-1}) 
\sigma(p_T) e^{\sigma^2(p_T)/4}}$ in eq. (\ref{bn}) depends on energy.
Using the above values for the boost invariant case in eq. 
(\ref{edb}) we find $\epsilon_0$=31 GeV/fm$^3$ with $T_{eff}$= 460 MeV
at RHIC and $\epsilon_0$=287 GeV/fm$^3$ with $T_{eff}$= 800 MeV at LHC.
The number density for the boost invariant case obtained from eq. 
(\ref{ndb}) is $n_0$=20/fm$^3$ at RHIC and 85/fm$^3$ at LHC.
Using the values of the $\sigma^2$ for boost non-invariant case we find
from eq. (\ref{ednb}): $\epsilon_0$=51 GeV/fm$^3$ with $T_{eff}$=521 MeV
at RHIC and $\epsilon_0$=517 GeV/fm$^3$ with 
$T_{eff}$=932 MeV at LHC for $\eta$=0.
The number density for the boost non-invariant case obtained from
eq. (\ref{ndnb}) is $n_0$=30/fm$^3$ at RHIC and 138/fm$^3$ at LHC for 
$\eta$=0. It can be noted that the energy density for the boost invariant
case is less than that of the boost non-invariant case. This is because
we have assumed a boost invariance in the entire rapidity range. 
However, the actual pQCD minijet rapidity distribution is not flat
in the entire rapidity range as can be seen from Fig. 2. Therefore,
in order to maintain the boost invariance in the entire rapidity range the 
density has to be smaller. However, in boost non-invariant case
the rapidity distribution of the minijet is minimaly altered and
the pQCD form of $\frac{dN}{dp_T dy}$ is used throughout the calculation.

Let us mention briefly other approaches for obtaining 
initial condition at RHIC and LHC. Commonly used way is to obtain 
energy density from eq. (\ref{edbj}) by using pQCD estimate of the
minijet $E_T$ distribution \cite{hijing,eskk,gwa,bhal}. This formula
assumes that the initial partons are uniformly distributed
in the coordinate space so that one divides the volume 
to obtain energy density. In \cite{nayak} a boost invariant Boltzmann
form $f(x,p,t_0)~=~e^{-p_T \cosh (\eta -y)/T}$ was used for the initial
distribution function with $\eta-y$ (boost invariant quantity)
correlation taken into account. In the parton cascade model \cite{gegm} 
momentum space structure function along with coordinate distribution
of the nucleons inside the nucleus was used directly in the transport
equation which is different from the minijet calculation by using pQCD.
In the approach of this paper we have obtained a phase-space
initial non-equilibrium distribution function $f(x,p)$ for the 
boost non-invariant
case with minimal altering the $p_T$ and $y$ distribution 
of the minijet partons obtained by using pQCD calculations.

To conclude, we have obtained a physically reasonable initial phase-space
distribution function of the partons formed at the very early stage of the
heavy-ion collisions at RHIC and LHC which can be used to study 
equilibration of the quark-gluon plasma \cite{bhal,nayak,cw}. 
To obtain such phase-space
distribution function we have considered several plausible parametrizations
of $f_i(x,p)$ and have fixed the parameters by matching
$\int f(x,p)p^\mu d \sigma_\mu$ to the invariant momentum space 
semi-hard parton distributions obtained using QCD (pQCD) as well
as fitting low order moments of the distribution. 
Once the parameters of $f(x,p)$ are found, we then have
determined the initial 
number density, energy density and the corresponding (effective)
temperature of the minijet plasma at RHIC and LHC energies. 
For a boost non-invariant minijet phase-space distribution
function we obtain $\sim$ 30(140) /fm$^3$ as the initial number density,
$\sim$ 50(520) GeV/fm$^3$ as the initial energy density and
$\sim$ 520(930) MeV as the corresponding initial effective 
temperature at RHIC(LHC). Our minijet initial conditions are obtained
above $p_T$=1(2) GeV at RHIC(LHC). As partons below these momentum
cut-off values might be described by the formation of a classical
chromofield, the $f(x,p,t_0)$ obtained in this paper is the 
distribution function for the matter part. Initial conditions
for lower transverse momentum has to be supplied in form of
the strength of the chromofield or in the form of field energy density
in the non-abelian transport equation to study production and
evolution of the quark-gluon plasma at RHIC and LHC. 

\acknowledgements
We thank Ulrich Heinz for useful discussions.


\begin{thebibliography}{aaaaaaaaaaaaaaa}

\bibitem{hijing} X. N. Wang, Phys. Rep. 280 (1997) 287;
X. N. Wang and Miklos Gyulassy, Phys. Rev. D 44
(1991) 1991; K. J. Eskola and K. Kajantie, Z. Phys. C 75 (1997) 515;
K. J. Eskola, K. Kajantie and J. Lindfors, Nucl. Phys. B
323 (1989) 37; I. Vitev and M. Gyulassy, Phys. Rev. Lett. (in press), 
hep-ph/0209161 and Talk given by I. Vitev at {\it hard probes in heavy 
ion collisions at LHC}, CERN, 7-11 October 2002.
\bibitem{larry1} 
L. M. A. Bettencourt, F. Cooper and K. Pao, Phys. Rev. Lett.
89 (2002) 112301, hep-ph/0109108;
L. Mclerran and R. Venugopalan, Phys. Rev. D 49 (1994) 2233,
Phys. Rev. D 49 (1994) 3352; Yu. Kovchegov and A. H. Mueller, Nucl. Phys.
B 529 (1998) 451; A. Kovner, L. McLerran and H. Weigert, Phys. Rev D
52 (1998) 3809, Phys. Rev. D 52 (1998) 6231.
\bibitem{others}
Y. Kluger, J.M. Eisenberg, B. Svetitsky, F. Cooper and E. Motolla,
Phys. Rev. Lett. 67 (1991) 2427; Phys. Rev. D45 (1992) 4659; 
F. Cooper, J. M. Eisenberg, Y.
Kluger, E. Motolla and B. Svetitsky, Phys. Rev. D 48 (1993) 190;
C. D. Roberts and S. M. Schmidt, Prog. Part. Nucl. Phys. 45
Suppl.1:1-103, 2000. 
\bibitem{nus}
A. Casher, H. Neuberger and S. Nussinov, Phys. Rev. D20 
(1979) 179; D. D. Dietrich, G. C. Nayak and W. Greiner, Phys. Rev. D64
(2001) 074006; hep-ph/0009178; J. Phys. G28 (2002) 2001; 
G.C. Nayak, D. D. Dietrich and W. Greiner, Rostock 2000/Trento
2001, Exploring Quark Matter 71-78, hep-ph/0104030;
G. C. Nayak and V. Ravishankar, Phys. Rev D55 (1997) 6877;
Phys. Rev. C58 (1998) 356.
\bibitem{al} A. H. Mueller, Nucl. Phys. B572 (2000) 227; K. J. Eskola
{\it et al.}, Nucl. Phys. B570 (2000) 379. 
\bibitem{grv98} M. Glueck, E. Reya and A. Vogt, Euro. Phys. J. C5 (1998) 461.
\bibitem{eks98} K. J. Eskola, V. J. Kolhinen and P. V. Ruuskanen, Nucl. Phys.
B 535 (1998) 351; K. J. Eskola, V. J. Kolhinen and C. A. Sagado,
Euro. Phys. J. C9 (1999) 61.
\bibitem{bjorken} J.D. Bjorken, Phys. Rev. D{\bf 27}, 140 (1983).
\bibitem{geig} 
E. Shuryak and L. Xiong, Phys. Rev. Lett. 48 (1982) 1066;
K. Geiger and J. Kapusta, Phys. Rev. Lett. 70 (1993) 1920; 
G. C. Nayak, Phys. Lett. B442 (1995) 427; X-M. Xu, D. Kharzeev, 
H. Satz and X-N. Wang, Phys. Rev. C53 (1996) 3051; G. C. Nayak, 
JHEP 9802 (1998) 005.
\bibitem{fred}  F. Cooper and  G. Frye,   Phys. Rev. D{\bf 10}, 186 (1974).
\bibitem{eskk}  K. J. Eskola {\it et al.}, Nucl. Phys. A696 (2001) 715; 
K. J. Eskola, V. J. Kolhinen, R. Vogt, Nucl. Phys. A696
(2001) 729 and references therein.
\bibitem{gwa} K. J. Eskola and M. Gyulassy, Phys. Rev. C47 (1993) 2329. 
\bibitem{bhal}
R.S. Bhalerao, G.C. Nayak, Phys. Rev. C 61 (2000) 054907.
\bibitem{nayak} G. C. Nayak, A. Dumitru, L. McLerran and W. Greiner, 
Nucl. Phys. A 687 (2001) 457.
\bibitem{gegm} K. Geiger and B. Muller, Nucl. Phys. B369 (1992) 600; 
K. Geiger, Phys. Rept. 258 (1995) 237.  
\bibitem{cw} C-W. Kao, G. C. Nayak and W. Greiner, Phys. Rev. D66 (2002)
034017; F. Cooper, C-W. Kao and G. C. Nayak, Phys. Rev. D66 (2002) 114016; 
hep-ph/0207370; J. Ruppert {\it et al.}, Phys. Lett. B 520 (2002) 233.

\end{thebibliography}
\end{document}